# Computational fluid dynamics modelling of microclimate for a vertical agrivoltaic system

Sebastian Zainali[1,*], Omar Qadir[1], Sertac Cem Parlak[1], Silvia Ma Lu[1], Anders Avelin[1], Bengt Stridh[1], Pietro Elia Campana[1,*]

[1]Mälardalen University. Dept. of Sustainable Energy Systems, Box 883, 721 23 Västerås (Sweden)

- Corresponding author:   sebastian.zainali@mdu.se

## Abstract

The increasing worldwide population leads to a constant increase in energy and food demand. These increasing demands have led to fierce land-use conflicts as we need agricultural land for food production while striving towards renewable energy systems such as large-scale solar photovoltaic (PV) systems, which also require in most of the cases agricultural flat land for implementation. It is therefore essential to identify the interrelationships between the food, and energy sectors and develop intelligent solutions to achieve global goals such as food and energy security. A technology that has shown promising potential in supporting food, and energy security, as well as support water security, is agrivoltaic (AV) systems. This technology combines conventional farm activities with PV systems on the same land. Understanding the microclimatic conditions in an AV system is essential for an accurate assessment of crop yield potential as well as for the energy performance of the PV systems. Nevertheless, the complex mechanisms governing the microclimatic conditions under agrivoltaic systems represent an underdeveloped research area. In this study, a computational fluid dynamics (CFD) model for a vertical AV system is developed and validated. The CFD model showed PV module temperature estimation errors in the order of 0-2 °C and ground temperature errors in the order of 0-1 °C. The shadings that occurred due to the vertical module had a reduction of the solar intensity of 38%. CFD modelling can be seen as a robust approach to analysing microclimatic parameters and assessing AV systems performances.

# 1 Introduction

The world's primary energy consumption ended at 173 Terawatt hours in 2019 [1]; however, this number is expected to double in the next mid-century due to the growth in the global economy and human population [2]. To meet this demand, the 17 Sustainable Development Goals (SDGs) cannot be overlooked. They represent an integrated framework of human, social and environmental development objectives [3] that provide a shared blueprint for peace and prosperity for the people and the planet. These goals are developed to tackle the following; climate crisis, ensuring that no one goes hungry, human rights abuses, and extreme poverty. In the past seven years, a tremendous progress towards this goals have been achieved. However, there is still a lot more work required [4]. Additionally, a result of the global growth in population and energy demand, more carbon emissions will be released into the atmosphere, which will significantly affect global warming and climate change. To tackle this carbon emissions increase, an agreement also known as the Paris Agreement was set to address the impacts and plan long-term goals of limiting global warming temperature to well below 2°C by long-term and sustainable adaptation [5]. In such context, the deployment of renewable energy resources (RES) will lead in meeting this demand as compared to non-renewable resources. Solar power is one of the most available and abundant renewable energies, and PV modules have lately become more affordable since the price of the modules has fallen by 10% every year for the last 30 years [6] . These are decisive factors that make the PV modules a key energy conversion technology.

Parallel to the growing energy demand, the demand of the agricultural sector will increase due to the increase of the global population and economy. Globally, agriculture accounts for approximately 70% of all freshwater withdrawals [7,8]. The scarcity of freshwater must be seriously addressed by promoting integrated water resources management with the aim of attaining the SDGs due to the close interrelationships between water, food and energy sectors [9]. Farmers and food production systems have succeeded in meeting the global food demand in recent times, but they will also face significant challenges to meet the future food demand which is expected to grow between 59% to 98% by 2050 [10]. To satisfy both agricultural and energy demands, large amounts of land must be used. However, along with water scarcity, worldwide we are also experiencing agricultural land scarcity due to the overuse of land for energy conversion (i.e., land used for growing energy crops), cities and networks expansion, climate changes, and desertification processes to cite some [11].

The Food, Energy, and Water (FEW) nexus is arguably one of the most essential approaches to reach sustainable development since it helps to better understand the complex interdependency between the FEW sectors and provides guidance on how to manage limited resources more sustainably. The framework is used to better understand the interrelationships and interdependencies between food, energy, and water. However, increased food, energy, and water demands together with climate change enlarge the conflict between the resources [12]. Nevertheless, the global community and representatives are solution-oriented and looking for new approaches to adapt according to climate change and development challenges [13]. One key technological innovation to overcome these challenges is combining agriculture and solar energy conversion are AV systems [14]. AV systems are a key technology in the FEW nexus perspectives, because of their advantage of reducing evapotranspiration and thus irrigation requirements and at the same time providing shading on crops, resulting in increased crop protection to adverse weather phenomena like droughts. Nevertheless, AV systems come with some major concerns, for instance, their impact on microclimatic conditions and crop productivity. Microclimate is defined as any climatic condition in a relatively small area within a few meters above and below the earth's surface including vegetation [15]. Conditions depend on, but are not limited to, factors such as humidity, temperature, wind, dew, frost, and evapotranspiration [16,17]. The research on this topic is however limited and the forces on which the microclimatic conditions depend, are not fully mapped yet [18]. This is especially true for the Nordic climate where extremely few AV systems are installed [19].

Biologists mostly credit light as the main driver for crop photosynthesis, although it is true for some crops such as sun crops that require high irradiation for full photosynthesis. However, shade-tolerant crops are saturated by low irradiation by the presence of higher chlorophyll content and higher antioxidant activity [20,21] . A group of Danish scientists investigated the effect of reduced light intensity on the growth of three common grass types, annual bluegrass, silky wind grass, and blackgrass [22]. The experiment was performed with six light levels aimed at 0%, 20%, 50%, 80%, 90%, and 95% shade corresponding to a mean daily light integral (DLI) of 12.4, 9.63, 7.13, 2.74, 0.95, and 0.69 mol.m$^{-2}$.d$^{-1}$. The DLI is the number of photosynthetically active photons accumulated in a square meter over the course of a day, and ranges in the photosynthetically active radiation (PAR) active region, it can be thought of as the crops daily "dose" of light for a given crop. The result of their experiment showed that a DLI between 0.69 to 3.71 mol.m$^{-2}$.d$^{-1}$ substantially reduced the height of the crops, the number of leaves, and stomatal conductance, it also delayed the flowering of wind grass and annual bluegrass, due to excessive shading [21]. Tanny et al. [23] also proved that complete shading as well as exposing crops to excessive radiation can also contribute to the stress factors, the latter process called photoinhibition. Consequently, the crops can respond by closing their stomata, thus, the tiny pores on the crops are not able to optimise the $CO_2$ uptake for photosynthesis. Fei et al. [23] investigated the potential of the japonica paddy rice variety (Heijing 5) for the Swedish climate. They investigated the rice grain in Uppsala for five consecutive years from 2013 to 2017. The polytunnel, a technique used to control greenhouse microclimate, was used for this project to cover the growing rice when the air temperature was forecast to fall below 10°C with a successful production. By producing favorable microclimatic conditions, they showed that rice could potentially be cultivated in countries with cold climatic conditions. Adeh et al. [24] investigated the impact sun irradiation had on soil moisture and pasture production. They found out that the open farmland tended to deplete the soil moisture more rapidly compared to partially and fully shaded AV systems since the water content in soil and crops are lost to the atmosphere due to evapotranspiration [24,25]. The findings concerning the reduction in evapotranspiration rates show the effectiveness of AV systems in areas marked out by high water scarcity as compared to open land farming [26,27]. Similarly, Marrou et al. [28] showed that the shade generated by the PV modules resulted in irrigation savings between 14-29%. Those results also show the potential to cultivate on land with a higher risk for drought.

Wind characteristics, i.e., wind speed and direction, influence the microclimatic conditions and the solar cells' performances. The lack of wind on sweltering days could potentially harm crop growth and the energy conversion efficiency of solar cells since the wind can provide a cooling effect [28,29]. On the other hand, strong wind speeds can significantly affect evapotranspiration rates [29,30]. AV systems can significantly affect wind characteristics depending on the layout. The crops´ water absorption is essential for photosynthesis. This phenomenon in turn is affected by the pressure in the atmosphere [31].

In numerous studies, CFD simulations have been used to analyse microclimatic impact [32–37]. CFD is used to solve complex flow equations with accurate numerical methods for both internal and external problems [38]. The interaction of solids and gases can be used to analyse the microclimate within agrivoltaic systems [34]. Baxevanou et al. [32] examined the effect of semi-transparent organic PV on a greenhouse roof cover by using CFD simulation to analyse the transmission radiation through the cover and into the greenhouse. Similarly, Majdoubi et al. [33] investigated the airflow circulation within a greenhouse and validated the model by temperature and relative humidity measurements from the crop fields. Additionally, CFD simulations have also been used for simulating PV system performances. For instance, Irtaza & Agarwal [37] investigated the turbulent wind effect on ground-mounted PV panels and found that the panels were vulnerable to high winds. Furthermore, Jubayer et al. [36] investigated wind-induced heat transfer from ground-mounted PV systems with reasonable prediction of mean wind speeds around the solar panel.

The microclimate study is of fundamental importance in understanding how the implementation of AV systems affects crop production and energy conversion. Therefore, in this study, a CFD model is developed to analyse and predict the microclimate of a vertical AV system. This study is an extension of a CFD PV model study by Johansson et al. [34]. The paper is structured as follows: Section 2 presents the studied system's input parameters and assumptions to develop the CFD model. Section 3 presents the validation results between measurements and the CFD model. Section 4 summarises the outcome of this study.

## 2 Method

### 2.1 Experimental AV system

The AV system investigated in this study is located at Kärrbo Prästgård in Västerås, Sweden (59.5549° N, 16.7585° E). As seen in Figure 1, the AV system is vertically-mounted with bifacial PV modules facing east and west. The AV system consists of 3 rows with a spacing of 10 meters to perform conventional agricultural practices.

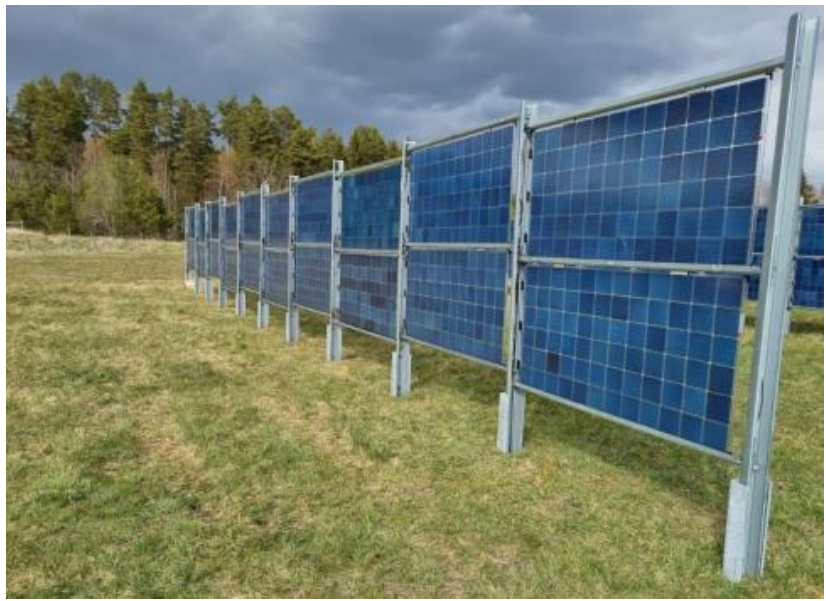

Figure 1 Vertical AV system at Kärrbo Prästgård.

## 2.2 Computational fluid dynamics model

The model representing the AV system is constructed in the computer-aided design (CAD) software Solidworks®. To perform CFD simulations, Solidworks Flow Simulation® which is embedded within Solidworks® CAD is used in this study. The modules are constructed with 5 layers and two aluminium poles. The assembly of the model is depicted in Figure 2. The layers used to construct the module are the following: 2 glass layers, 2 EVA plastic layers, and 1 cell layer.

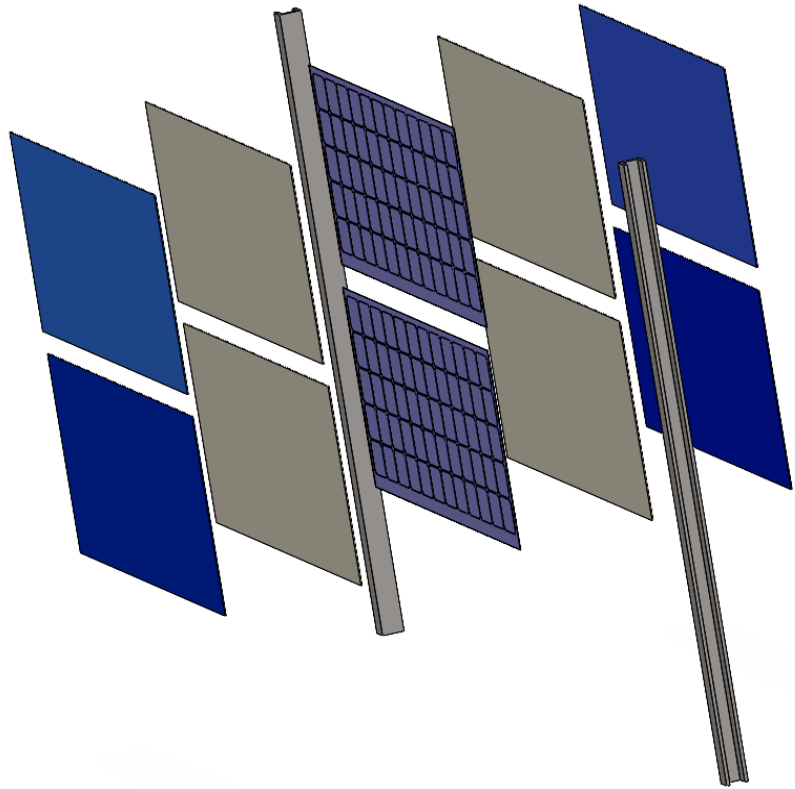

Figure 2 PV module assembly.

The representing model of the reference plant consist of 30 module assemblies attached to the ground as depicted in Figure *3*. The dimensions of the design used in this study are attached in Table 1 below.

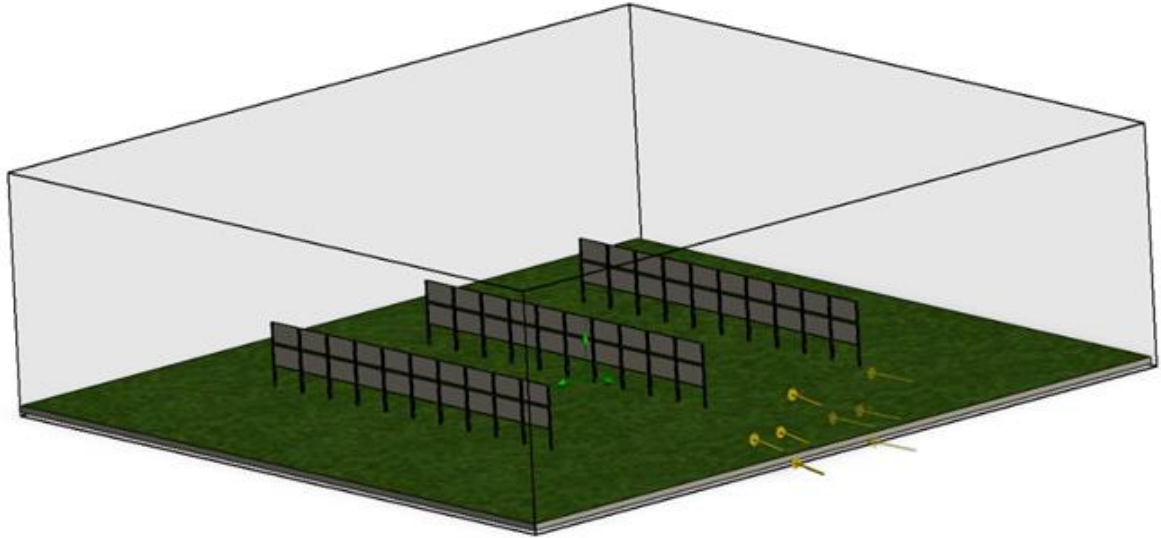

Figure 3 AV system assembly

Table 1 Agrivoltaic system dimensions.

| Part | Height mm | Width mm | Thickness |
|---|---|---|---|
| m-Si Cell | 157 | 157 | 0.25 |
| EVA Plastic | 900 | 2000 | 1 |
| Glass | 900 | 2000 | 2,5 |
| Pillar | 2800 | 100 | 80 |
| m-SI Cell layer | 900 | 2000 | 0,5 |
| Soil | 40 000 | 15000 | 300 |

Solidworks flow simulation® generates a global mesh for the entire computational domain. The global mesh can be refined by increasing the number of levels to produce more fine cells but it will take more central processing unit (CPU) time and require more computer memory. In Solidworks flow simulation® the level of initial mesh is scaled from 1 to 7, and in our study the level is set to 6. The model is set up as a transient model to perform analysis for each specific time step. The computer used for simulation had an advanced micro devices (AMD) Ryzen 9 5900X 12-Core Processor 3.70 GHz with 32 GB of random access memory (RAM) and 1 TB Solid-state Drive (SSD). The material's thermal properties used in this study can be seen in Table 2 below.

Table 2 Material characteristics.

| Materials/Gas | Specific Heat Capacity J/kg⁰C | Density kg/m³ | Thermal Conductivity W/m⁰K |
|---|---|---|---|
| Air (20⁰C) | 1006.5 | 1.2 | 0,029 |
| m-Si Cell | 703.98 | 703.98 | 149 |
| EVA-Plastic | 1900 | 930 | 0.23 |
| Glass | 800 | 2500 | 0.98 |
| Pillar | 420 | 7900 | 45 |
| Soil | 1600 | 1600 | 2 |

## 2.3 Validation

The CFD model is validated with data gathered on the 23$^{rd}$ of June 2022 with a temporal resolution of 5 min. The measured ambient temperature, global horizontal radiation, diffuse horizontal radiation, and wind rose are shown in Figure *4* and Figure *5*. The maximum ambient temperature was around 26 °C at 15:00, while the minimum temperature of 13 °C was during the early morning around 03:00. There were some clouds during the morning between 06:00-09:00, which can be seen on the global and diffuse horizontal irradiance. The rest of the day was mostly clear sky and the maximum global horizontal irradiance recorded was around 900 Wm$^{-2}$. The wind rose shows that most of the wind recorded had a velocity between 0-8 ms$^{-1}$ and the wind direction was southwest during the whole day.

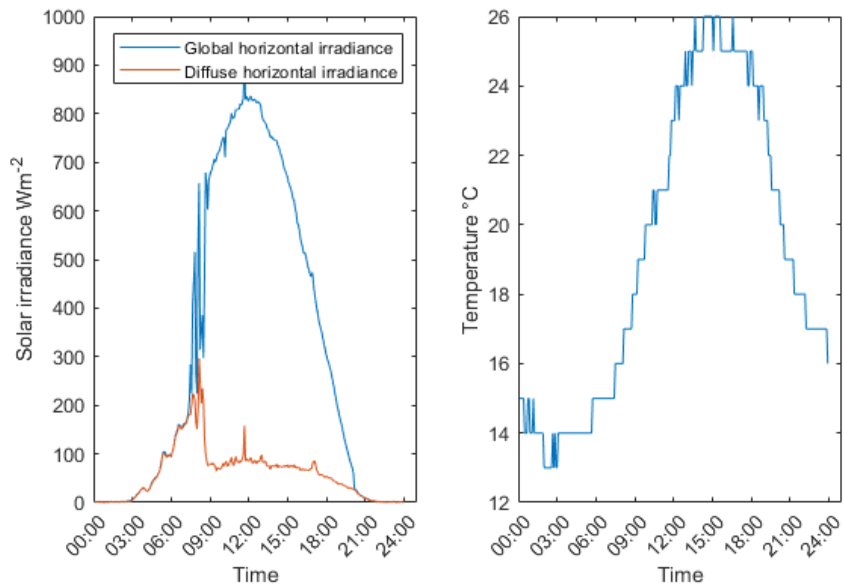

Figure 4: Solar radiation (left) and ambient temperature (right) at Kärrbo Prästgård 23rd of June 2022.

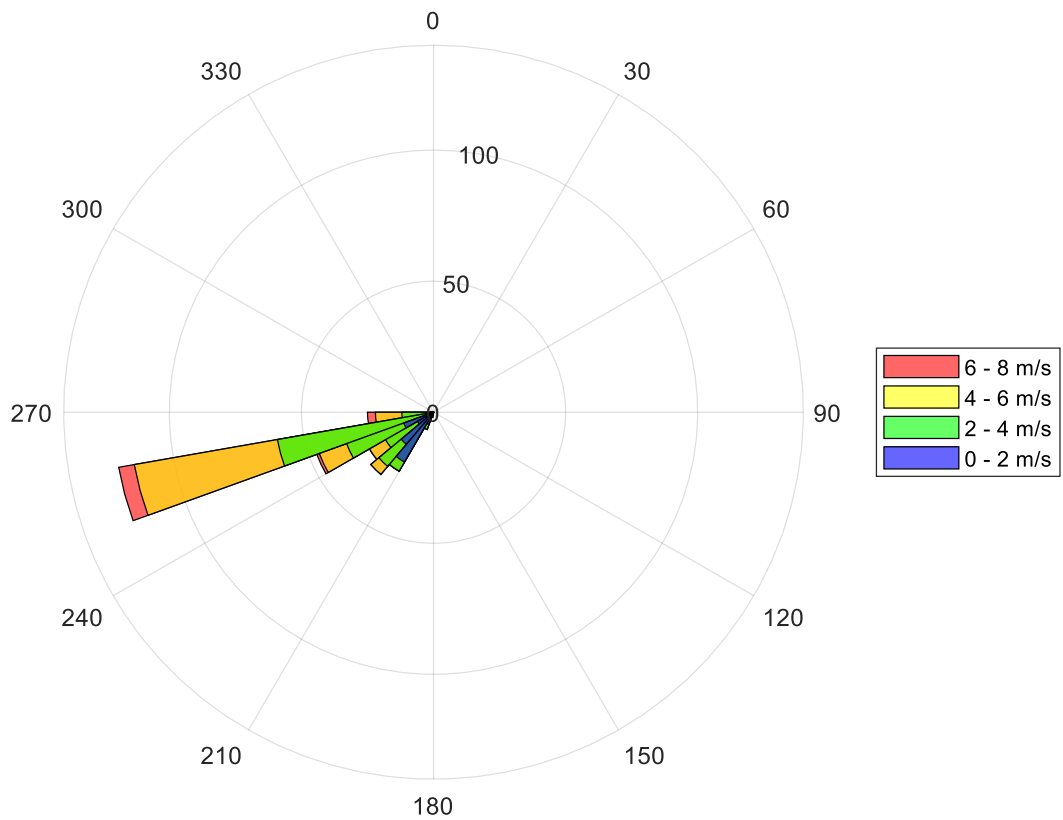

Figure 5: Wind rose measured at Kärrbo Prästgård on the 23rd of June 2022.

The model validation is performed by using 4 different sensors installed on the field while a thermal camera is used to measure the temperature of the bifacial modules. The characteristics of the sensors are summarized in Table 3, including the thermal camera.

Table 3 Sensor specifications.

| Sensor | Specifications |
|---|---|
| Solar-Log Sensor Box Professional Plus | Radiation range: 0 W/m$^2$ to 1400 W/m$^2$ |
| | Radiation accuracy: 5 W/m$^2$ ± 2.5% (0 W/m$^2$ to 1400 W/m$^2$) |
| | Ambient temperature range: -40 °C to +90 °C |
| | Wind speed range: 0.9 m/s to 40 m/s |
| | Cell temperature range: -40 °C to +85 °C |
| | Cell temperature accuracy: ± 1 °C |
| Lufft WS600-UMB Smart Weather Sensor | Temperature range: -50 °C to +60 °C |
| | Temperature accuracy: ±0.2 °C (-20 °C to 50 °C) |
| | Wind direction range: 0 ° to 359.9 ° |
| | Wind direction accuracy: <3 °, RMSE > 1.0 m/s |
| | Wind speed range: 0 m/s to 75 m/s |
| | Wind speed accuracy: ±0.3 m/s (0 m/s to 35 m/s) |
| Solar Survey 200R | Temperature range: -30 °C to 125 °C |
| | Temperature accuracy: 1° |
| Delta-T SPN1 Sunshine Pyranometer | Radiation range: 0 W/m$^2$ to > 2000 W/m$^2$ |
| | Radiation accuracy: ±5% ±10 W/m$^2$ Hourly averages |
| | Temperature range: -40 °C to +70 °C |
| | Temperature coefficient: 0.02% per °C |
| | Spectral response: 400 nm to 2700 nm |
| | Spectral sensitivity variation: 10% |
| FLIR i50 thermal camera | Temperature range: -20 °C to +350 °C |
| | Accuracy: ±2% |
| | Spectral range: 7.5μm to 13μm |

The temperature measurement on the Solar-Log sensor boxes is not fully representative of a bifacial module due to their difference in thermal characteristics. The Solar-Log sensors are monofacial solar cells and do not consider the contribution from the rear side radiation affecting the thermal balance. Therefore, cross-validation with a simplified model presented in Leonardi et al. [39] is used to calculate the bifacial PV module temperature $T_{bi}$ (°C) and is given by

$$T_{bi} = \frac{T_{front} + T_{rear}}{2}, \qquad (1)$$

where $T_{front}$ and $T_{rear}$ are the front and rear temperatures of the bifacial PV module. The model presented in Leonardi et al. [39] was one of the most performing model among those compared in Johansson et al. [34]. The module temperature obtained from the CFD simulation is validated by comparing it with the simplified model and thermal images. The ambient temperature, wind speed, and wind direction used as inputs in the CFD model are gathered from the Lufft WS600-UMB smart weather sensor. The solar radiation used as input to the CFD model is gathered from the Delta-T SPN1 sunshine pyranometer. The incident solar radiation on the modules is compared with the vertical east and west facing Solar-Log sensor boxes measuring both solar radiation and panel temperature. The ground temperature is compared with Solar Survey 200R temperature probe. The temperature probe is positioned in the middle of the row with a depth of 10 cm.

## 3 Results and discussion

This section first presents the validation results by comparing the CFD model with measurement data from the experimental agrivoltaic system. The section is then followed by presenting the microclimatic variation within an agrivoltaic system.

### 3.1 Model validation

The validation of the CFD model for the incident solar irradiance on the PV modules is presented in Figure 6. The CFD model underestimates the incident solar irradiance compared to the Solar-Log sensor boxes measurements. During cloudy weather conditions, it could be noticed that there is a significant difference in solar irradiance at some occurrences between the CFD model and the east face solar cell. This can be attributed to the accuracy of the decomposition model implemented in the CFD software. Overall, the intensities are very similar between the CFD model and the Solar-Log sensor boxes. However, it could be seen that the CFD model does not reach the same maximum intensity as the Solar-Log sensors.

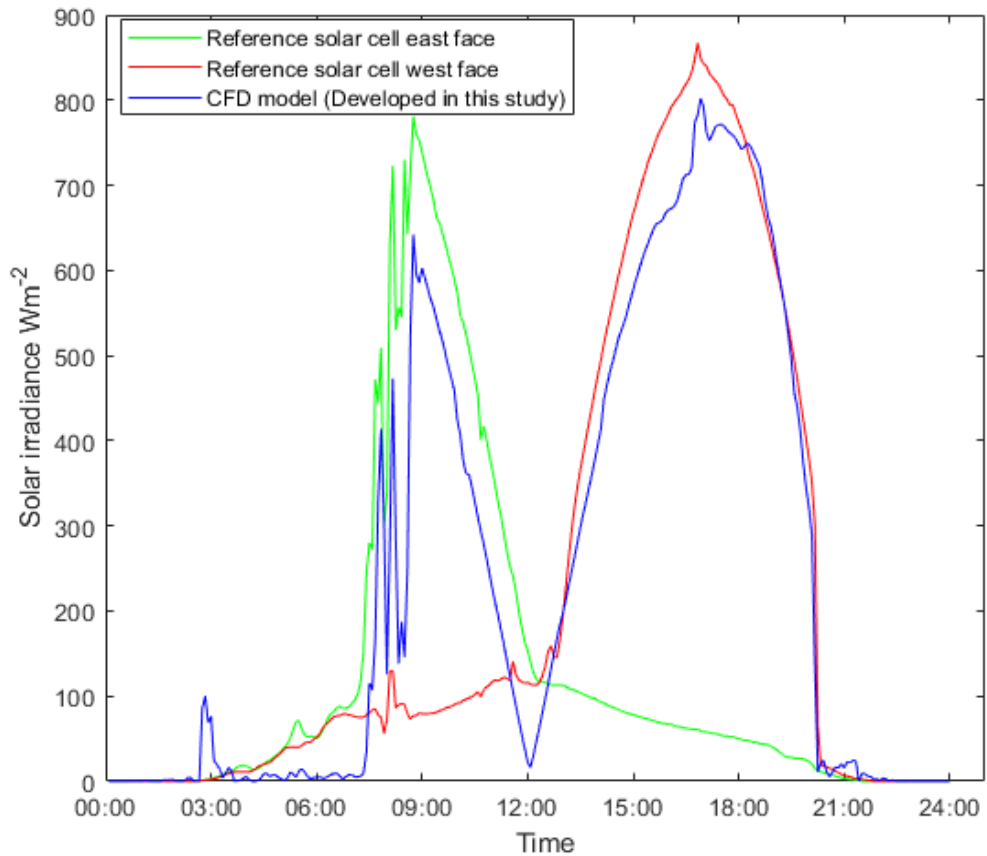

Figure 6 Measured solar irradiance distribution at Kärrbo Prästgård 23$^{rd}$ of June 2022 and calculated solar irradiance from the CFD model.

The incident solar irradiance in the CFD model at 17:00 local time is presented in Figure *7*. The incident solar irradiance is higher on the panels than on the ground as the maximum incident solar irradiation is obtained when an object is perpendicular to the sun's rays. In Figure *7*, the PV modules receive above 800 Wm$^{-2}$ while the ground receives less than 400 Wm$^{-2}$. It could also be noticed that the shadings from the PV modules further reduce the intensity at the shaded areas to less than 250 Wm$^{-2}$. The shaded areas have a 38% reduction in solar intensity compared to non-shaded areas. The distribution of light reaching the crops' is important to better understand the impact of AV systems. As observed by Yasin et al. [22], a reduction of light affects certain crops growth significantly. The CFD analysis can improve our understanding of light reduction for specific AV system designs and how crop growth is affected.

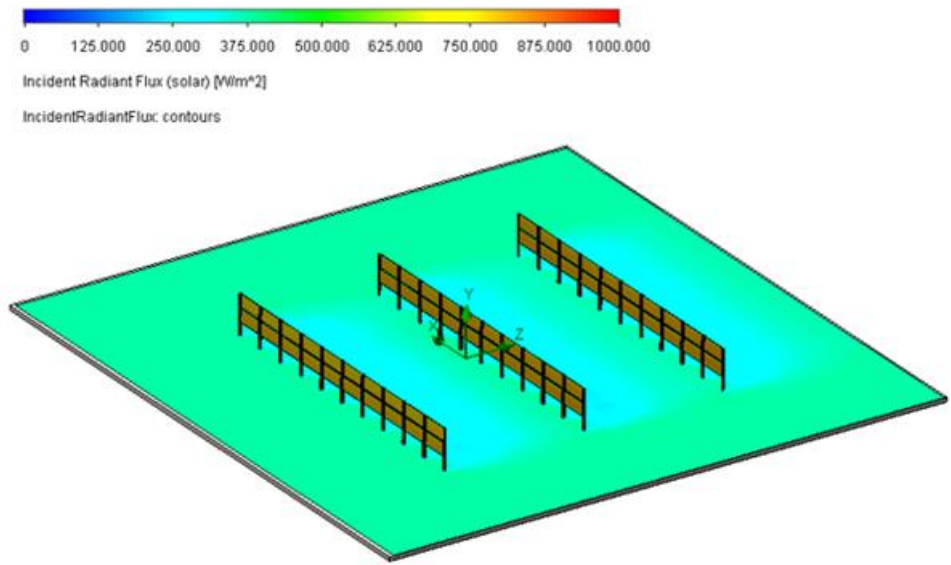

Figure 7 Computed CFD model solar irradiance at 17:00 local time of the 23$^{rd}$ of June 2022

The solar irradiance affects the temperature on the ground and PV modules. In Figure *8* and Figure *9*, the CFD model temperature validation is presented. The CFD model had a temperature ranging from 14 °C to 38 °C. The reference solar cell east face (Solar-Log sensor) reached temperatures of about 40 °C, significantly higher than all other methods. The thermal camera had a similar temperature range to the CFD model, which was between 12 °C and 37 °C. The CFD model had estimation errors of 0-2 °C compared to the thermal camera readings. Figure *10* shows thermal camera pictures for 11:00, 14:00, and 16:00. The Leonardi et al. [32] model had some overestimations during the morning compared to the thermal camera similar to the CFD model and the reference solar cell east face. However, during the hours when the sun reaches the highest intensity, it could be seen that the Leonardi et al. [32] model mainly underestimates due to the low-temperature readings from the Solar-Log sensor on the rear side of the PV module these underestimations was also observed by Johansson et al. [34], when analysing a vertical bifacial PV module in Solidworks®.

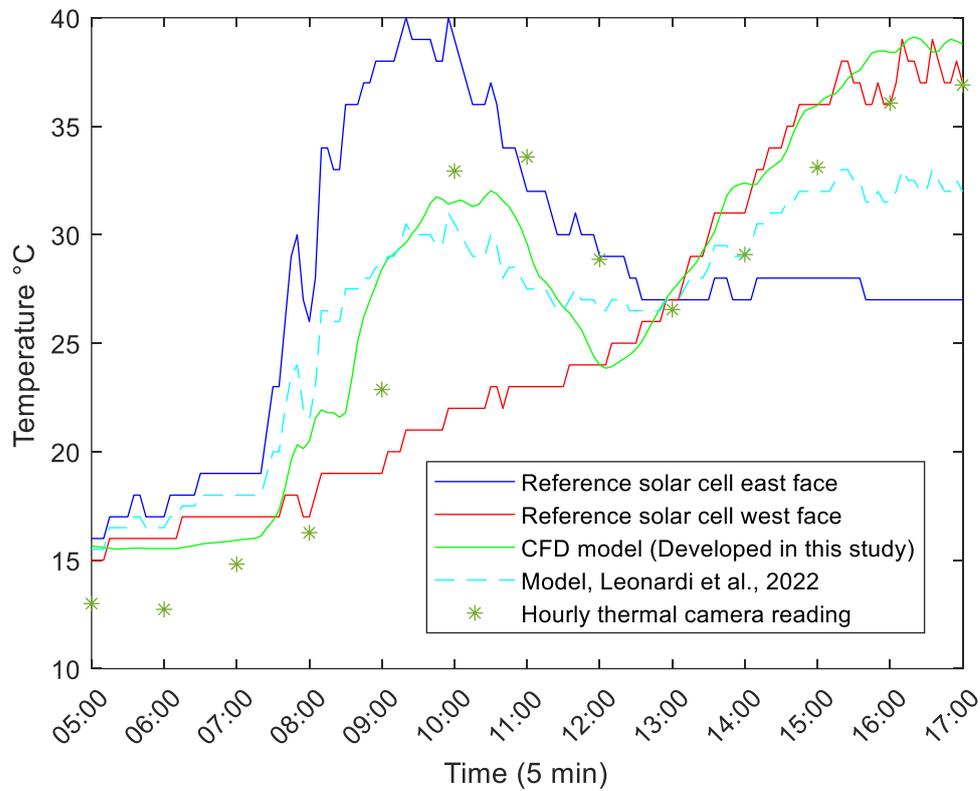

Figure 8 Comparison between measured PV Module temperature at Kärrbo Prästgård 23rd of June 2022, and PV modules temperature computed with Leonardi's model [39], thermal camera readings, and PV modules temperature calculated with the CFD model developed in this study.

The ground temperature in the CFD model has a temperature error of less than 1 °C compared to the readings from the Solar Survey 200R. The CFD ground temperature range was between 17 °C to 22 °C. The ground has very high inertia and does not fluctuate similarly to the PV module temperature. However, this could be problematic during heat waves as the temperature will remain high before decreasing, increasing the stress on plants. This phenomenon could be further analysed with the CFD simulation model but is out of the scope of this work.

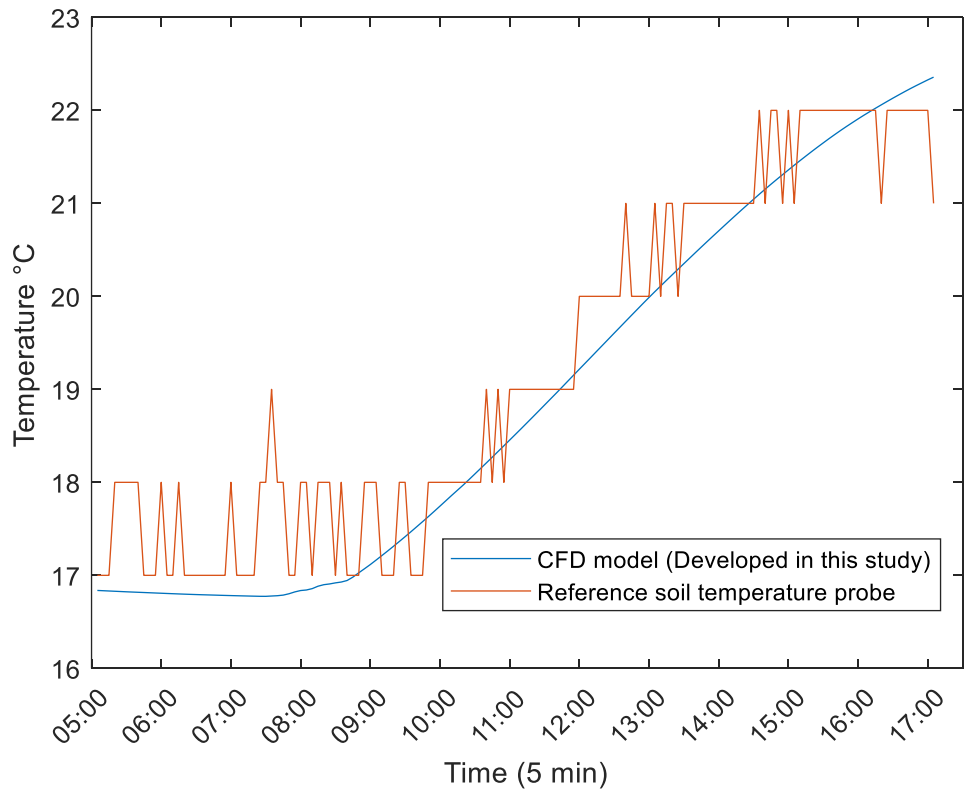

Figure 9 Comparison between measured ground temperature at Kärrbo Prästgård 23rd of June 2022 and ground temperature calculated with the CFD model developed in this study.

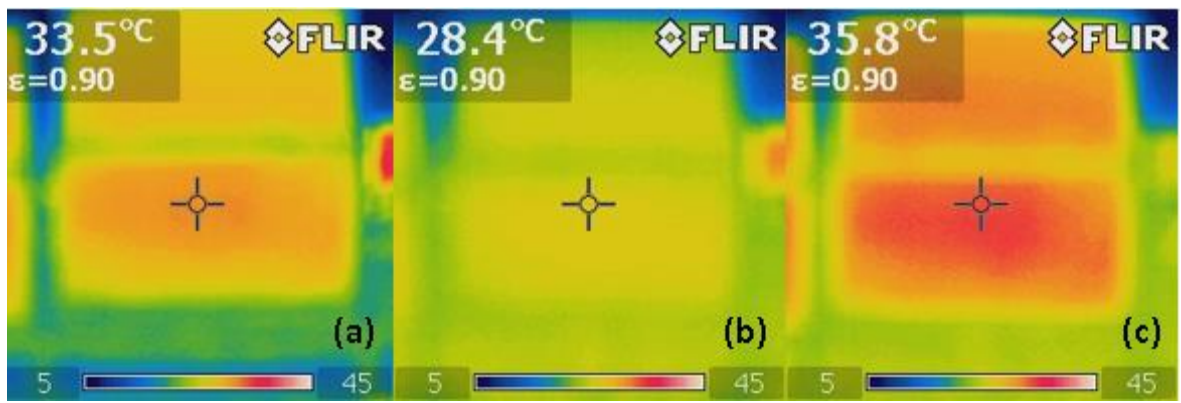

Figure 10 Thermal camera readings with a temperature scale between 5 °C to 45 °C at 11:00 (a), 14:00 (b), and 16:00 (c).

### 3.2 Microclimatic variation within an agrivoltaic system

Using the CFD model, it is also possible to analyse other microclimate parameters for an AV system, such as wind directions, velocities, and pressures. In Figure *11*, the wind speed and directions are presented at 17:00 local time. The wind direction is from the west and wind speed varies between 4 to 8 ms$^{-1}$, where it could be noticed that the wind speed decreased at ground heights. A slight difference could also be seen between the incoming and outgoing wind speeds, where the outgoing wind speed reaches above 7 ms$^{-1}$.

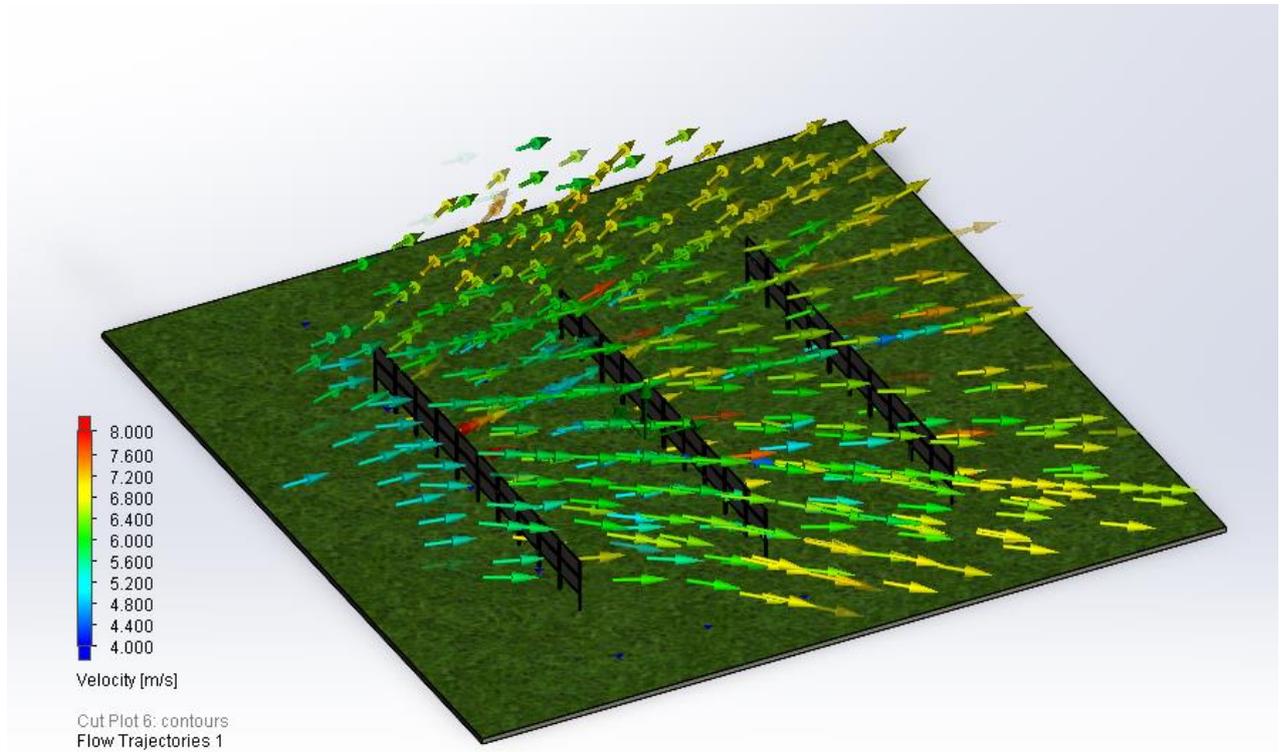

Figure 11 CFD model wind speed and direction at 17:00 local time.

It can be observed that the wind speed profiles in Figure *11* and ambient temperature in Figure *12* change within the AV system. A similar observation was found by Marrou et al. [28]. To what extent the changes in wind speed profiles and ambient temperature can be due to numerous sources of variation within the AV system as described by Marrou et al. [28]. The air temperature changes significantly within an AV system where the heat from the PV modules are increasing the air temperature within the AV rows as observed in Figure *12*. The air temperature varies from 23.5°C to 23.8°C at 17:00 local time and the highest temperature can be observed at the PV modules. The increased air temperature from PV modules is an already known phenomenon where the air temperature increase can influence the climatology for large-scale solar installations with a temperature increase of up to 3-4°C [40]. Armstrong et al. [41], recorded an air temperature increase up to 5.2°C under the modules in a solar park. However, these recorded air temperatures are not of the same magnitude as noted in the CFD simulation.

The observed temperature increase in the CFD simulation was 0.3°C at 17:00 local time. Nevertheless, Hassanpour et al. [24] measured a significant difference in mean temperature close to the PV panels in an elevated agrivoltaic system. However, their measurements are of the same magnitude as the CFD simulation air temperature and varied between 0-0.3°C.

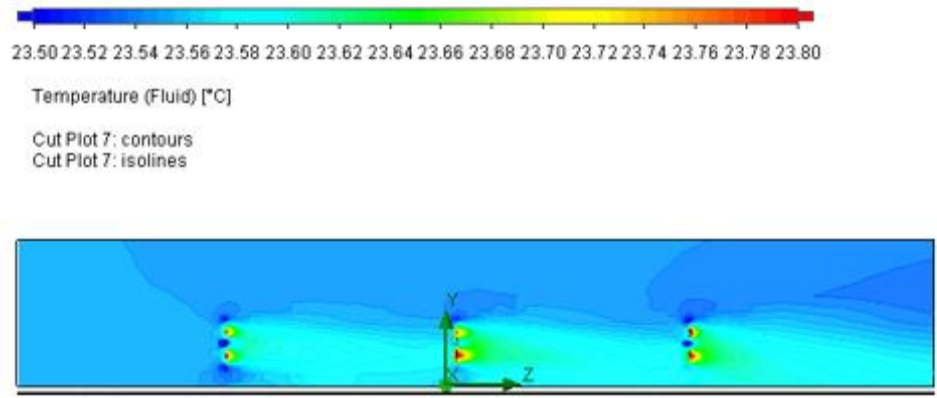

Figure 12 CFD model air temperature between each row at 17:00 local time.

The CFD model developed and validated in this study shows how a CFD approach could be implemented to analyse microclimatic conditions under AV systems. The knowledge of microclimatic conditions is essential to estimate better the effects on crop yield and energy conversion in AV systems. The vertically mounted AV system show that microclimatic conditions can significantly change throughout the AV system. It is therefore essential to analyse and validate in further studies how the microclimate conditions vary in larger-scale and different types of AV systems.

## 4 Conclusions

In this study, a 3D CFD model of a vertically mounted AV system is developed to analyse the microclimatic conditions. and the CFD model is validated with measurements performed at an experimental AV system installed at Kärrbo Prästgård, Västerås, Sweden. The main conclusions of this study are:

- The incident solar irradiance estimates from the CFD model are close to the measured incident solar irradiance. The CFD model generally underestimates the peak solar irradiance, especially in cloudy conditions.
- The incident solar irradiance can vary significantly on the ground for a vertical AV system. The shadings produced by the vertical PV modules can reduce the solar intensity by 38% compared to non-shaded areas on the investigated day. This reduction in irradiance hitting the ground surface can influence energy balances and temperatures accordingly.

- The PV module temperature in the CFD model has shown a good agreement with thermal camera readings. The CFD model had an estimation error of 0-2 °C compared to the thermal camera readings. The ground temperature in the CFD model had a temperature error of less than 1 °C compared to the measurements.

- The microclimate varies significantly throughout an AV system. CFD modelling can analyse several essential microclimate parameters to understand better the variations that occur by implementing AV systems.

# 5    Acknowledgments

The authors acknowledge the financial support received from the Swedish Energy Agency through the project "SOLVE solar energy research center", grant number 52693-1. The main author also acknowledges the financial support received from European Energy Sverige AB. The authors also acknowledge the financial support received from the Swedish Energy Agency through the project "Evaluation of the first agrivoltaic system in Sweden", grant number 51000-1. The author Pietro Elia Campana acknowledges Formas - a Swedish Research Council for Sustainable Development, for the funding received through the early career project "Avoiding conflicts between the sustainable development goals through agro-photovoltaic systems", grant number FR-2021/0005. This article is based on the results of the master thesis from Omar Qadir and Sertac Cem Parlak [42].